\input psfig
\documentstyle[twoside,fleqn,espcrc2]{article}

\newcommand{\AmS}{{\protect\the\textfont2
  A\kern-.1667em\lower.5ex\hbox{M}\kern-.125emS}}

\title{ QED2 as a testbed for interpolations between
 quenched and full QCD}
\author{W.~Bardeen\address{Fermilab, P.O. Box 500, Batavia, IL 60510},%
  A.~Duncan\address{Dept. of Physics and Astronomy,University of Pittsburgh,
 Pittsburgh, PA 15260},%
  E.~Eichten$^{a}$%
  and 
  H.~Thacker\address{Dept. of Physics, University of Virginia, 
 Charlottesville, VA 22901}}
\begin{document}
\begin{abstract}
 Lattice  QED2 with the Wilson formulation of fermions is used as a convenient
 model system to study artifacts
 of the quenched approximation on a finite lattice. The quenched functional
 integral is shown to be ill-defined in this system as a consequence of the
 appearance of exactly real modes for physical values of the fermion mass.
 The location and frequency of such modes is studied as a function of
 lattice spacing, lattice volume, topological charge and improved action
 parameters. The efficacy of the recently proposed modified quenched 
 approximation is examined, as well as a new approach to the interpolation
 from the quenched to full dynamical theory employing a truncated form
 of the fermion determinant.
 
\end{abstract}

\maketitle
\section{Introduction}
 In this talk, some general features of the Wilson-Dirac spectrum in
 quenched lattice gauge theory are discussed using  2-dimensional
 QED as a convenient model system \cite{QEDpaper}.  The specific focus will be the
 dependence of the real part of the spectrum on the parameters of the
 theory. The nonexistence of the quenched functional integral is 
 found to arise from a complicated analytic structure induced by these
 real modes. The relation of quenched, pole-shifted \cite{QCDpaper} and full dynamical
 amplitudes is also discussed. Finally, the usefulness, accuracy and feasibility 
 of an interpolating determinant approach to the full theory can be studied
 in detail in this model. 

\section{General Features of the Wilson-Dirac spectrum}
 In QED2 quark 
 propagators are inverses of a matrix $D-rW+m\equiv {\cal M}+m$, with $D$, $W$ and $m$ the
 naive Dirac matrix, $W$ the Wilson term, and $m$ a quark mass parameter:

\begin{eqnarray}
\label{eq:Mdef}
 {\cal M}&\equiv& D-rW \\
\label{eq:Ddef}
 D_{a\vec{m},b\vec{n}}&=&\frac{1}{2}(\gamma_{\mu})_{ab}U_{\vec{m}\mu}\delta_{\vec{n},\vec{m}+\hat{\mu}} \nonumber \\
 &-&\frac{1}{2}(\gamma_{\mu})_{ab}U^{\dagger}_
{\vec{n}\mu}\delta_{\vec{n},\vec{m}-\hat{\mu}} \\
\label{eq:Wdef}
 W_{a\vec{m},b\vec{n}}&=&\frac{1}{2}\delta_{ab}(U_{\vec{m}\mu}\delta_{\vec{n},\vec{m}+\hat{\mu}}+U^{\dagger}_{\vec{n}\mu}\delta_{\vec{n},\vec{m}-\hat{\mu}} )
\end{eqnarray}
where $a,b$ are Dirac indices, $\vec{m},\vec{n}$ lattice sites, $U$ the unimodular link
 variables, and the Wilson parameter $r$ is usually taken to be unity. Quite a lot is known 
 about the spectrum of $ {\cal M}$, which is complex as $W$ is hermitian while $D$ is
 skew-hermitian:\\

(1) The norm of the quadratic form $ {\cal M}$ is less than or equal to 2 for arbitrary
 gauge fields \cite{Gattringer}, so the spectrum is contained inside a circle of radius 2
 in the complex plane. In
 fact, a typical spectrum (see Fig. 1) has an elliptical shape with four critical branches, two in
 the center and one on either side. Conventionally the left critical branch 
 represents the chiral (zero fermion mass) limit.\\
  (2)  The secular polynomial for $ {\cal M}$ has real coefficients and only even terms, so
 eigenvalues necessarily appear as real doublets $\lambda,-\lambda$ or as complex
 quartets $\lambda,\lambda^{*},-\lambda,-\lambda^{*}$. In particular, the appearance
 of exactly real eigenvalues (despite the fact that $ {\cal M}$ is not a normal matrix) is
 generic, and such eigenvalues persist in finite neighborhoods of any gauge configuration
 point with a real mode. (Note the 2 exactly zero modes associated with each critical
 branch in Fig. 1).\\
(3) The appearance of exactly real modes for $-2 \leq \lambda \leq -\frac{1}{2\kappa_{c}}$
 (i.e. for physical naive fermion masses using the left critical branch) will lead to 
 nonintegrable singularities in the quenched functional integral involving lattice Wilson-Dirac
 propagators. The integral can be defined by analytical continuation from the nonsingular
 region $|\lambda| >2$, but the region inside the spectral ellipse is thoroughly infested
 with complicated branch cuts connecting a large number of branch points. A pinch
 argument shows that such branch points arise at any eigenvalue of  $ {\cal M}$ for
 gauge configurations where the link variables $U$ are either +1 or -1. The noisy 
 behavior of quenched simulations can be  traced directly to this pathology.

\begin{figure}
\psfig{figure=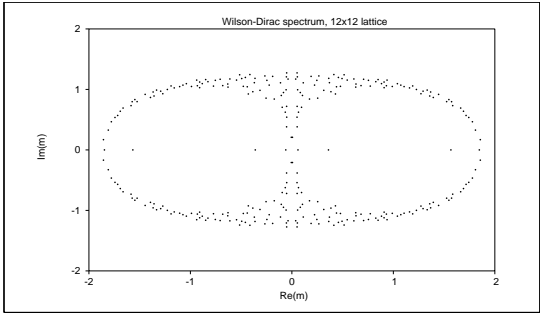,
width=0.95\hsize}
\vspace{-0.4in}
\caption{A typical Wilson-Dirac spectrum in QED2}
\label{fig:spectrum}
\end{figure}

 The above statements are analytically demonstrable, but even more can be learned from
 detailed explicit simulations. For example:\\

(4)  The integer part of the topological charge $Q_{1} \equiv\frac{1}{2\pi}\sum_{P}\sin{(\theta_{P})}$, (where  $\theta_{P}$ is the plaquette angle for plaquette $P$), tracks quite closely the number of
 exactly zero modes per critical branch. Transitions between different topological
 charge sectors in the course of the simulation are accompanied by movement of
 complex eigenvalue quartets towards and then along the real axis.\\
(5)  Histograms of the exactly real modes accumulated over many  (typically 1000) 
 decorrelated configurations for different beta values, but keeping the physical
 lattice volume fixed show that the spread of real modes into the physical mass
 region becomes acute at strong coupling, and that the probability at fixed
 physical fermion mass  of encountering
 exceptional configurations in which a nearby real propagator pole introduces large
 fluctuations in measured hadronic amplitudes decreases rapidly as
 beta is increased (see Fig. 2).\\
(6)  With increasing lattice volume at fixed $\beta$, the 
  probability of encountering an exceptional configuration as one approaches the 
  left crtical line decreases with increasing volume if one keeps a fixed offset
 from the critical line to maintain a fixed physical quark mass.  However, exceptional configurations necessarily appear at any volume
 once one goes sufficiently close to the critical point.  \\
(7)  The frequency and distribution of exactly real modes 
  is not 
 substantially affected by a clover improved action. Of course, on any
 individual configuration, the location of real modes (if present) will change with the value
 of the clover coefficient chosen. But the statistical noise introduced by exceptionals 
 in any large ensemble remains.

\begin{figure}
\psfig{figure=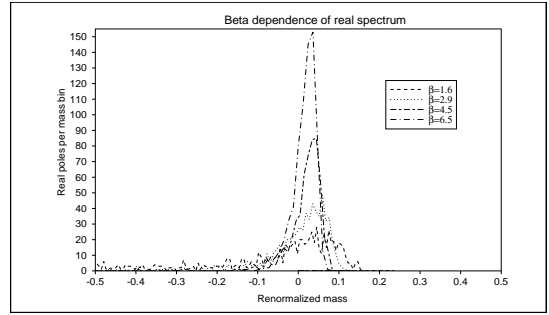,
width=0.95\hsize}
\vspace{-0.35in}
\caption{Beta dependence of  histogram of real modes}
\label{fig:betahistograms}
\end{figure}

\section{Comparison of Quenched, MQA and Full Dynamical Simulations}
  Recently, we have proposed  
 a modified quenched approximation (MQA) in which 
 the quenched functional integral is made well-defined by a pole-shifting procedure
 which incorporates the correct spectral behavior in the continuum limit (see 
 \cite{Estitalk} for a more  detailed description). 
QED2 offers a convenient model for comparison
 of naive quenched, MQA and full dynamical results. A typical result is shown 
 in Fig. 3, where the pseudoscalar correlator (``pion propagator") is shown at a
 bare quark mass of 0.08 (at $\beta$=4.5, 10x10 lattice) for these 3 cases. The
 statistical noise in the quenched correlators is essentially eliminated in the MQA
 results, which also are found to interpolate between the naive quenched and
 full dynamical results. This is gratifying- the MQA, in addition to rendering 
 quenched amplitudes meaningful on coarse lattices, appears to move us closer
 to the unquenched theory.

\begin{figure}[htp]
\vspace{-0.4in}
$$\psfig{figure=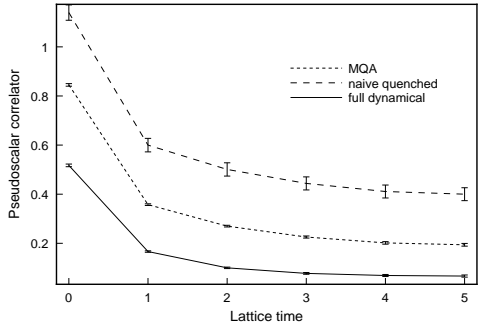,
width=0.90\hsize}$$
\vspace{-0.6in}
\caption{Pseudoscalar correlators in QED2- quenched, MQA and full dynamical}
\end{figure}

\vspace{-0.4in} 

\section{Interpolating Determinant Approach to Dynamical Fermions}
 We have recently begun a study of an alternative approach to the problem of
 interpolating between  quenched and unquenched gauge theory, inspired by
 the insights gained in the MQA work on the role of small eigenvalues. The idea is
 to separate off  and include explicitly  in the simulation the infrared contributions 
 to the determinant. In superrenormalizable QED2, the lowest  $2N_{\lambda}$ eigenvalues 
 contribute essentially all of the fluctuations to $\ln {\rm det}(\gamma_{5}({\cal M}-m))$,
 as indicated in Fig.4, while the remaining 200-$2N_{\lambda}$ (on a 10x10 lattice)
 hardly contribute to the determinantal variation. As a consequence correlators 
 computed using just the lowest 10\% of the spectrum are essentially exact (see Fig.5).
 
\begin{figure}[htp]
\vspace{-0.3in}
$$\psfig{figure=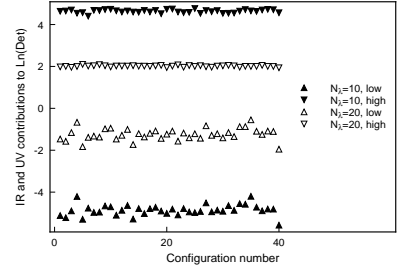,
width=0.74\hsize}$$
\vspace{-0.5in}
\caption{Fluctuations in Log(Det) from low and high eigenvalues (dynamical simulation
at $\beta$=4.5)}
\end{figure}
\begin{figure}[htp]
\vspace{-0.3in}
$$\psfig{figure=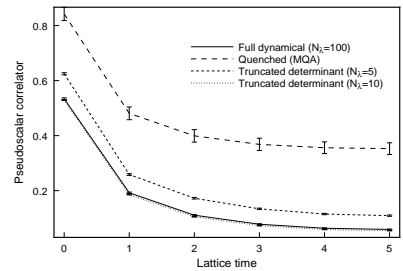,
width=0.74\hsize}$$
\vspace{-0.5in}
\caption{Comparison of quenched, truncated determinant and unquenched
correlators}
\end{figure}

 In QCD4 the UV part of the quark spectrum certainly contributes importantly to a
 renormalization of coupling, visible as a substantial shift of scale in lattice amplitudes.
 However, work in progress shows that all the important infrared physics (e.g.
 the correct chiral structure, eliminating quenched chiral logs),  say up to a scale
 of 300 MeV ,  can be built in by
 inclusion of  a few hundred eigenvalues of  $\gamma_{5}({\cal M}-m)$ which are readily
 accessible by a Lanczos scheme \cite{Kalk}.  It seems possible that the remaining determinant
 effects not simply reducible to a change of scale may be included at the end by a reweighting scheme, or perhaps by using an appropriate loop representation \cite{Sext} for the intermediate  part
 of the quark spectrum.  A study of these issues in QCD4 is in progress.

\vspace{-0.1in}


\end{document}